# Electromagnetic Crimping on Threaded Surface: FEM Modelling, Validation and Effects of Pitch and Discharge Energy on Deformation in an Empirical Relation

Deepak Kumar [a, *], Shafeeque E. S.[a], Sachin D. Kore [b], Arup Nandy [a]

[a] Department of Mechanical Engineering, Indian Institute of Technology Guwahati, Guwahati and 781039, India

[b] School of Mechanical Science, Indian Institute of Technology Goa, Goa and 403401, India

*Corresponding author's email address: kumar176103003@iitg.ac.in

## Abstract

Electromagnetic crimping is a high-velocity joining method to join highly conductive workpieces where a pulsed magnetic field is applied without any working medium or mechanical contact to deform the workpiece. This work explores tube-to-tube joining of Copper outer tube and Stainless steel threaded inner tube using electromagnetic crimping. A non-coupled simulation model is developed for the finite element analysis. ANSYS Maxwell package is used to obtain the magnetic field intensity, which is later converted to pressure using an analytical equation, and this pressure is applied to the two-tube working domain in ANSYS Explicit Dynamics. Numerical simulations are done for different combinations of discharge energies and pitches of the thread to analyse deformation, stress and strain. The converged finite element results are validated using experimental data. The amount of deformation is found to be proportional to discharge energy and the pitch of the thread used. An empirical relation is developed for the deformation as a function of discharge energy and pitch. The relation is able to predict the deformation for other discharge energies, which is later verified with ANSYS simulations.

## 1  Introduction

Multi-material assemblies have come up as the need of the hour due to various benefits over conventional engineering materials like weight reduction, better resistance to corrosion, enhanced strength, and electrical conductivity. There is further necessity to optimise these material designs for efficient production with better ergonomics and reduced polluting discharges. Copper-Steel is such a dissimilar material assembly with numerous applications in the electrical and power sectors, as discussed by Patra et al. [1] and Simoen et al. [2]. Due to copper's limited solubility in iron, the conventional joining of Cu-Steel might lead to hot cracking [3]. Therefore, solid-state joining methods such as Electromagnetic welding (EMW) are recommended to join dissimilar metals. Magnetic pulse welding further offers higher speed, precise control and is able to form overlap joints in sheets and axisymmetric tubes even with dissimilar materials [4]. Electromagnetic crimping (EMC), an electromagnetic forming process variant, makes use of a high-speed pressure pulse to deform the material [5]. The rapid discharge of current from a charged capacitor bank creates a high-intensity magnetic field in the crimping area. This magnetic field produces a radial Lorentz force that deforms the workpiece without any physical contact between them. The gap between them depends on the active area or the crimping zone [6].

Lai et al. have developed a coupled analytical model consisting of an electromagnetic actuator and a uniform pressure coil for analysis of the forming process [7]. Yu et al. have done a sequentially coupled field analysis of compression of tubes by electromagnetic forming using the ANSYS software package [8]. After coupled finite element model simulation for tubular joints using compression welding, Khan et al. have proposed an effective weldability window for Cu-Al tube to rod joint by predicting the magnetic field and magnetic pressure distribution [9]. Shanthala et al. have simulated the electromagnetic welding of mild steel tubes to steel bars using COMSOL MULTIPHYSICS software and validated them using experimental results [10]. Pawar et al. have performed a loose coupled analysis of electromagnetic forming of a muffler tube using ANSYS MAXWELL and ANSYS EXPLICIT DYNAMICS [11]. Kumar et al. have performed finite element analysis in LS-DYNA$^{TM}$ of Cu-SS tube-to-tube joint with a smooth interface and validated it with experimental results along with a novel finite element model to predict the pull-out strength of the joint [12] [13]. After considering the groove geometry, material properties and the electrical characteristics of the forming equipment, Weddeling et al. proposed an analytical model for determining the process parameters such as the discharge energy [14]. Chaharmiri et al. have studied how magnetic pressure and resulting radial displacement is affected by a stepped field shaper during EM forming by simulation using ANSYS MAXWELL and ABAQUS software package followed by experimental validation [15]. Kumar et al. have performed a comparison between the non-coupled simulation model (MAXWELL and ANSYS EXPLICIT DYNAMICS) and the coupled (LS-DYNA $^{TM}$) simulation model of electromagnetic crimping [16]. Pawar et al. have studied the coupled and non-coupled simulation models for the electromagnetic perforation of tubes using the same combination of software packages [17].

According to Weddeling et al. (2011), the EMC joints can be categorised into three types: interference-fit joints, form-fit joints and joints using adhesive bonds. This is based on which mechanism dominates when the joint is subjected to external load [18]. Tubular workpieces can be joined by any combination of the three or all three of them [5]. The occurrence of interference fit joint is by the formation of interference pressure between the workpieces after deformation, and it is based on the difference in the elastic recovery between two joined workpieces according to Mori et al. [19]. Form-fit joints are made by inserting one flyer workpiece material into a groove by making an undercut on the target workpiece. Thus, at the macroscopic level, a mechanical interlock is formed, and the strength of this interlock determines the resistance against axial loads. Christian et al. have studied how groove geometries influence the strength of aluminium form-fit joints while varying the shape and size of the grooves [18]. Kumar et al. have analysed how surface profile affects a Cu-SS interference-fit tube joint by conducting pull-out tests, compression tests and torsion tests and concludes that knurled joints provide the best joint strength compared to threaded or smooth surfaces [20]. Design parameters of threads are found to be easier to vary and optimise than knurled joints.

Electromagnetic crimping has a high potential for applications in the automotive [21] [22], electrical [23], nuclear [24] and aerospace [25] industries. Therefore, analytical and numerical modelling and hence the prediction of the properties of the joint can be significant. In this work, a copper (outer) tube and a stainless steel (inner) tube with a threaded outer surface are joined using the electromagnetic crimping technique to form a tube-to-tube joint. Numerical simulations are performed using ANSYS MAXWELL and ANSYS EXPLICIT DYNAMICS. A non-coupled model is used, so the magnetic field data obtained from MAXWELL is converted to corresponding magnetic pressure data using an analytical equation. This pressure load is applied to the tube to tube joint in ANSYS Explicit Dynamics, and the resulting deformation is validated with experiments. An empirical relation is derived from the experimental data to predict the deformation of the tube for other intermediate process parameters.

## 2 Working Principle of electromagnetic crimping

The power source used here is a charged capacitor bank. It discharges a damped sinusoidal current, which acts as the primary current and generates a magnetic field. By Faraday's law of induction, a secondary eddy current is generated in the field-shaper by this magnetic field. A field-shaper or a field-concentrator, as patented by Khimenko et al. [26], is usually an axisymmetric element built using a high electrically conductive material [5]. The eddy current in the field-shaper leads to a secondary magnetic field which in turn produces a tertiary induced current in the outer tube's outer surface. The magnetic field generated by this tertiary current interacts with the secondary magnetic field and produces a Lorentz force. This force causes elastic deformation in the inner SS tube and plastic deformation in the outer copper tube [16]. These variants in deformations and subsequent elastic recovery of the outer tube and inner tube lead to the formation of an interference-fit joint. The fitting of the outer tube inside the threads of the inner tube causes a form-fit joint. If the electric field intensity is denoted by $\vec{E}$, the magnetic flux density is denoted by $\vec{B}$, and the time is denoted by t, then Maxwell's equation governs the electromagnetic process as

$$\vec{\nabla} \times \vec{E} = \frac{-d\vec{B}}{dt} \qquad (1)$$

If the total current density is denoted by $\vec{J}$ and electrical permittivity is denoted by $\varepsilon$,

$$\vec{\nabla} \times \vec{H} = \vec{J} + \varepsilon \frac{d\vec{E}}{dt} \qquad (2)$$

$$\vec{\nabla} \bullet \vec{B} = 0 \qquad (3)$$

If total charge density is denoted by $\rho$,

$$\vec{\nabla} \bullet \vec{E} = \frac{\rho}{\varepsilon} \qquad (4)$$

If the source current density is denoted by $\vec{J}_S$ and electrical conductivity is denoted by σ, from Ohm's Law, we have

$$\vec{J} = \sigma \vec{E} + \vec{J}_S \qquad (5)$$

If the vector potential is denoted by $\vec{A}$, from Eq.(3), magnetic flux density is written as,

$$\vec{B} = \vec{\nabla} \times \vec{A} \qquad (6)$$

If the magnetic permeability of the medium is denoted by $\mu_m$, from Eq.(6), the total current density $\vec{J}$ induced is written as,

$$\vec{J} = \vec{\nabla} \times \frac{\vec{B}}{\mu_m} \qquad (7)$$

$$\vec{J} = \vec{\nabla} \times \frac{1}{\mu_m}\left(\vec{\nabla} \times \vec{A}\right) \qquad (8)$$

Using Maxwell's equation, the acting body force in the form of Lorentz force, $\vec{F}$ is written as,

$$\vec{F} = \vec{J} \times \vec{B} = \left(\vec{\nabla} \times \frac{\vec{B}}{\mu_m}\right) \times \vec{B} \qquad (9)$$

Magnetic pressure $p$ acting on the workpiece surface is determined by integrating the force per unit volume over the thickness of the tube. If the radius, thickness, the penetrated magnetic field and the magnetic field in the gap between the tubes are denoted by $r$, $t'$, $H_{pen}$, $H_{gap}$,

$$p(r,t') = \int_{r_0}^{r_i} F(r,t')dr = \frac{1}{2}\mu\left(H_{gap}^2(t') - H_{pen}^2(t')\right) \qquad (10)$$

If the thickness is more than the skin depth, the infiltrated magnetic field can be generally ignored so that magnetic pressure is given as,

$$p(t') = \frac{1}{2}\mu H_{gap}^2(t') \qquad (11)$$

## 3  Experimental Setup

The electromagnetic crimping experiments are conducted on an electromagnetic forming machine. Table 1 shows the material properties for Copper and SS304.

Table *2* shows the circuit parameters of the machine. A circular cross-sectioned wire is used to make a 13-turn copper solenoid coil of length 91 mm and an inner diameter of 49, as shown in Figure 1. When the coil is shorter than the tube, the magnetic field distribution is non-homogeneous, and an equal length of coil and workpiece results in a uniform magnetic field, as reported by Yu et al. (2007) [27]. Pawar et al. have studied the EM forming of muffler tube and has observed that equal length coil produces a uniform magnetic field and maximum deformation compared to all other longer and shorter coil lengths [28]. While analysing the electromagnetic compression process, the variation of homogeneous radial deformation with the variation of tube to coil length ratio has been studied by Yu et al. (2010), and an optimum ratio has been determined [29].

Table 1 Material Properties of Copper and Stainless Steel

| Properties | Electrical conductivity (S/m) | Ultimate tensile strength (MPa) | Young's modulus (GPa) | Poisson's Ratio |
|---|---|---|---|---|
| **Copper** | $59.6 \times 10^6$ | 251 | 130 | 0.34 |
| **SS-304** | $1.389 \times 10^6$ | 505 | 193 | 0.29 |

The detailed dimension of the solenoid coil, inner SS304 and outer copper tubes and the single-step field-shaper are shown in Figure 1. A copper tube of cross-section Ø12.7 × 0.9 mm thick is used as the outer tube. A SS 304 tube of Ø10 × 2 mm thick and with threaded surface is used as the inner tube. Table 3 and Table 4 (Kumar et al., 2021) [12] shows the chemical composition of Copper and SS304 respectively.

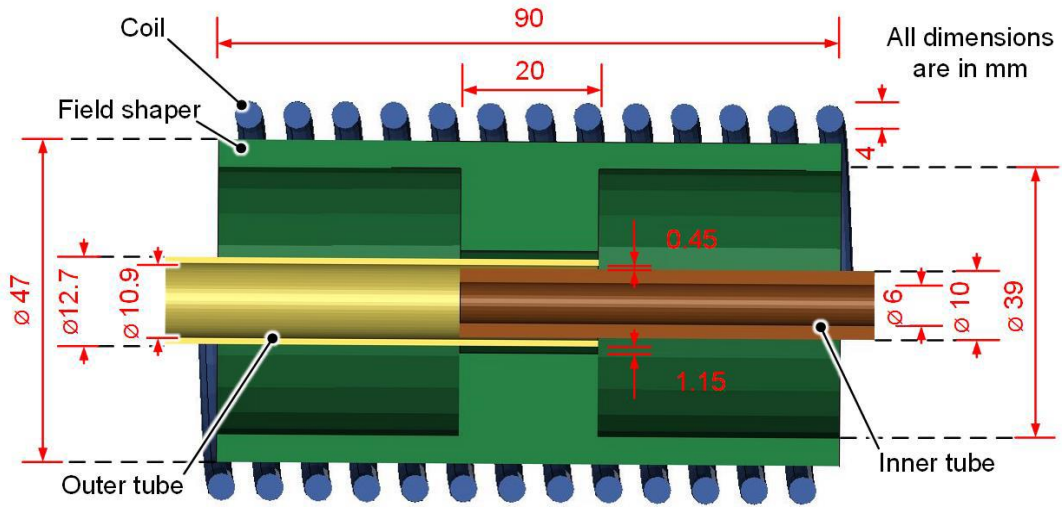

Figure 1 Detailed Dimensioning of the coil, tubes and field-shaper

Table 2 Circuit parameters

| Frequency | Maximum voltage | Maximum energy | Circuit resistance (R) | Capacitance (C) | Circuit inductance (L) |
|---|---|---|---|---|---|
| 20 kHz | 15 kV | 10 kJ | 12.5 m Ω | 90 µF | 0.7 mH |

Electromagnetic crimping experiments are performed at three different discharge energy values: 3.4kJ, 3.9 kJ, 4.4 kJ. A Rogowski coil and an oscilloscope are used to measure the current waveforms at three discharge energy values.

## 4 Finite element modelling

The flow chart for the non-coupled simulation model is shown in Figure 2.

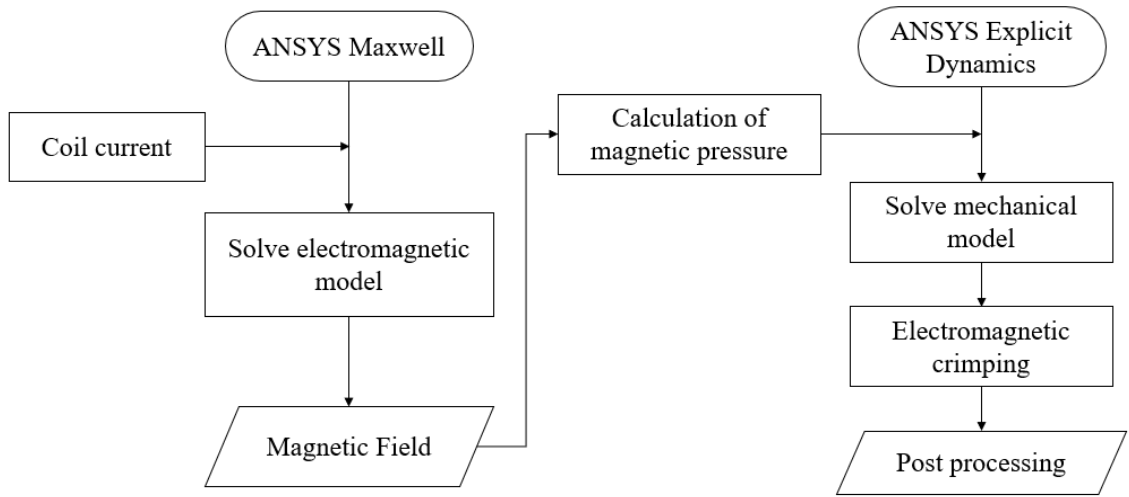

Figure 2 Flow Diagram for non-coupled simulation

ANSYS Maxwell is an electromagnetic field simulation software which uses finite element method to solve time-varying electric and magnetic field. It uses an adaptive meshing technique. We use ANSYS Maxwell to solve for the electromagnetic problem. The length of both tubes is 100 mm. The model developed in ANSYS Maxwell is shown in Figure 3.

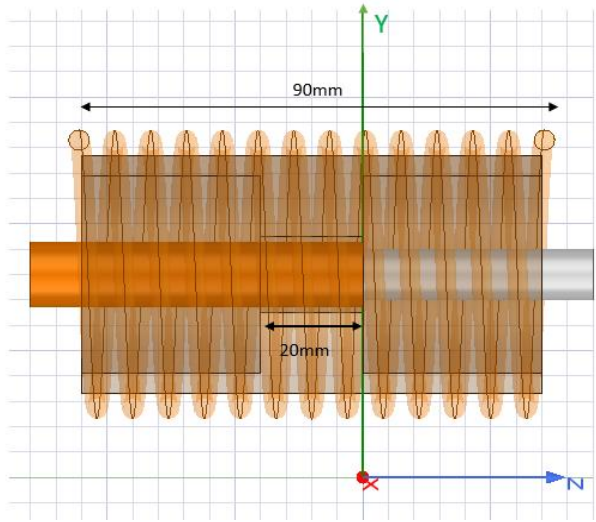

Figure 3 Model developed in ANSYS Maxwell

The copper cable used in the coil is 4 mm in diameter, and the pitch of the coil is 7 mm. The coil contains 13 turns and is of length 100 mm. There are input and output terminals at either end of the coil. Terminals are also made of copper wire of length 25 mm and diameter 4mm. At the input terminal of the coil, current is applied as shown in Figure 4. The end time for the simulation is $3 \times 10^{-5}$ s.

The following assumptions are made in this simulation model:

- The current distribution in the coil cross-section is uniform.
- The electrical properties of the material, such as conductance and permeability, are assumed to be constant throughout the time period.
- Joule heating loss has not been considered.

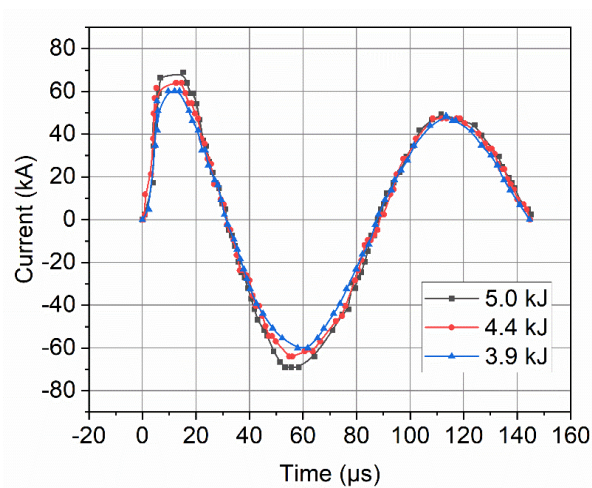

Figure 4 First pulse of the current waveform in the coil for three discharge energies

ANSYS Maxwell solves the resultant magnetic field, and this magnetic field intensity is converted to corresponding mechanical pressure using the equation (11).

ANSYS Explicit Dynamics is a multi-physics software capable of solving short-duration, structural problems having complex contact interactions with or without friction, which may include linear and non-linear buckling, fracture and fatigue and more [30]. The problems may have nonlinearities in the geometry, such as large deformation and large strain, including rate-independent and rate-dependent plasticity. They make use of both linear and non-linear material models. Structural analysis having very small time steps, such as high-speed impacts or explosions, can be performed efficiently using the Explicit Dynamics package.

The schematic of the model developed in ANSYS Explicit Dynamics is shown in Figure 5(a). The ends of the tubes are fixed using boundary conditions in ANSYS to replicate the experimental setup. The front view and the sectional views are shown in Figure 5(b) and Figure 5(c), respectively. The pressure data obtained from Maxwell is used here as input. The equivalent pressure is applied uniformly along the length in the crimping zone. Convergence study is used to determine the optimum mesh size and time step.

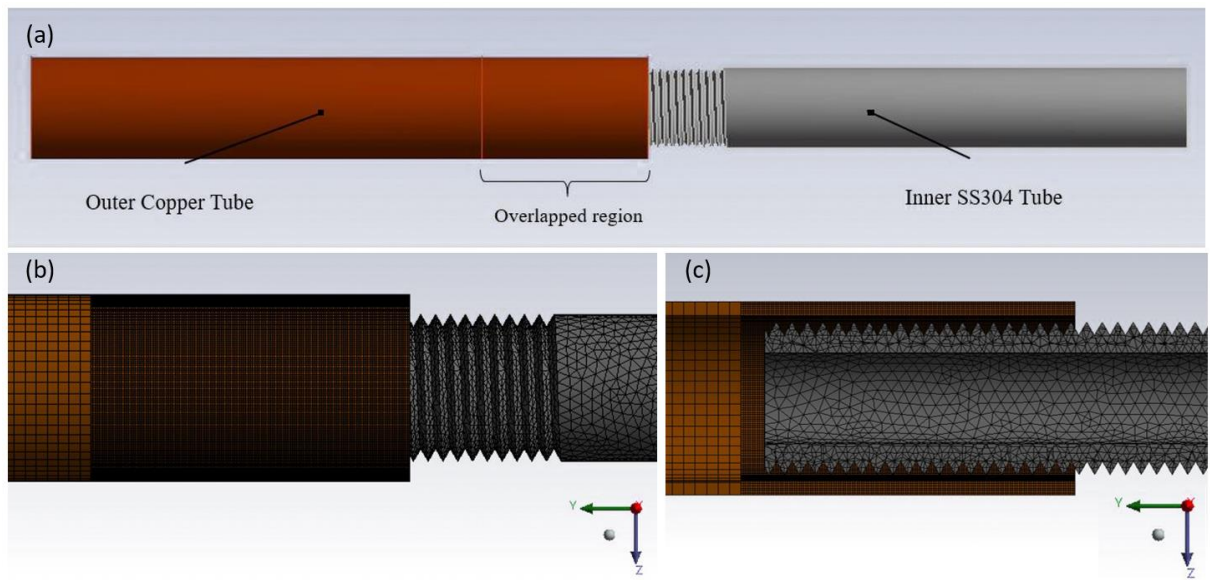

Figure 5 (a) Outer Copper tube and Inner SS304 tube (b) Model with mesh (c) Sectional view with mesh

## 4.1 Material Model

### 4.1.1 Johnson-Cook (J-C) Model

J-C material model is recommended for high temperature and high strain-rate processes. So, it is used for both the Copper and SS304 workpieces. If the equivalent plastic stress is denoted by $\sigma_y$, equivalent plastic strain, equivalent plastic strain rate and reference equivalent plastic strain rate are denoted by $\varepsilon$, $\dot{\varepsilon}$, and $\dot{\varepsilon}_0$ respectively, the material constants are denoted by A, B, C, n, m and absolute temperature, melting temperature and room temperature are denoted by $T$, $T_m$ and $T_r$, the J-C model can be described as,

$$\sigma_y = \left(A + B\varepsilon^n\right)\left(1 + C\ln\left(\frac{\dot{\varepsilon}}{\dot{\varepsilon}_0}\right)\right)\left(1 - \left(\frac{T - T_r}{T_m - T_r}\right)^m\right) \tag{12}$$

Parameters of the J-C material model for Cu and SS 304 are shown in Table 3. Properties of the materials are considered to be isotropic for this simulation.

Table 3 Constants of J-C material model for Cu and Stainless Steel [5] [31]

| Material | n | A(MPa) | B(MPa) | $T_m$ (K) | m | C |
|---|---|---|---|---|---|---|
| Copper | 0.31 | 90 | 292 | 1331 | 1.09 | 0.025 |
| SS 304 | 0.36 | 350 | 275 | 1723 | 1 | 0.022 |

## 5 Results and Discussion

### 5.1 Convergence analysis in ANSYS Maxwell with mesh size refinement

Mesh refinement is done at a constant time step and for three different number of total elements. All the other input parameters are kept constant. The discharge energy used is 3.9 kJ. From Figure 6, it is observed that the maximum magnetic field (H) value increases with time, reaches a maximum at $1.4 \times 10^{-5}$ s and then decreases. Along the length, the magnetic field is maximum between 35-55 mm (crimping zone). This intensification and concentration of magnetic field in the crimping zone is achieved using the field shaper. The solution converges at a mesh with 5000 elements.

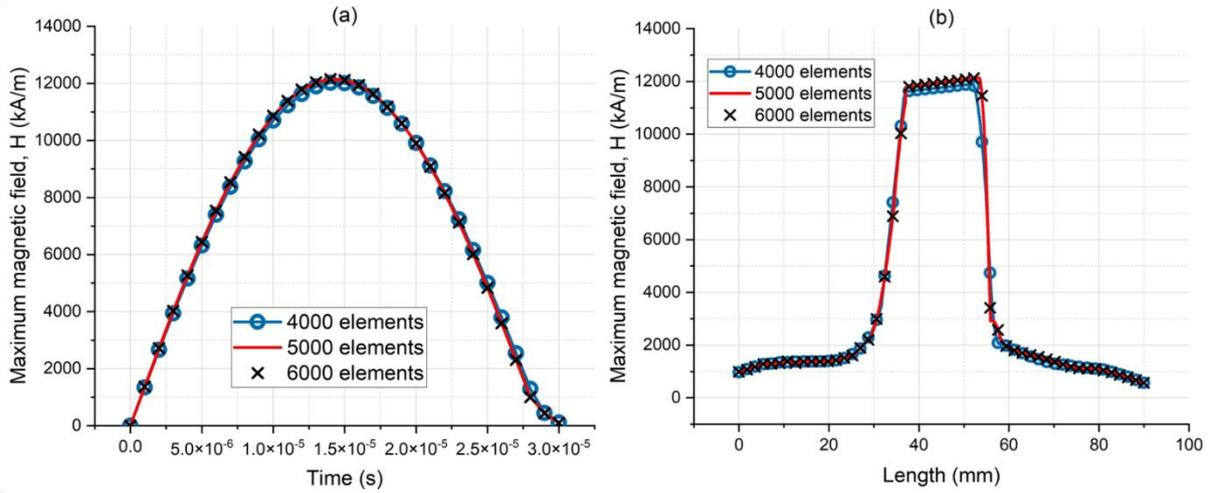

Figure 6 (a) Variation of H with time and (b) variation of H along the length at time $1.4 \times 10^{-5}$ s

Timestep refinement is automatically controlled as per the refined mesh by ANSYS Maxwell, and the optimum time step is chosen.

## 5.2 Calculation of magnetic flux density and pressure for different discharge energies

The magnetic field data is converted to corresponding magnetic pressure by using the equation (11). The maximum magnetic field at time $1.4 \times 10^{-5}$ s and corresponding maximum pressure variation are plotted in Figure 7 (a) and (b).

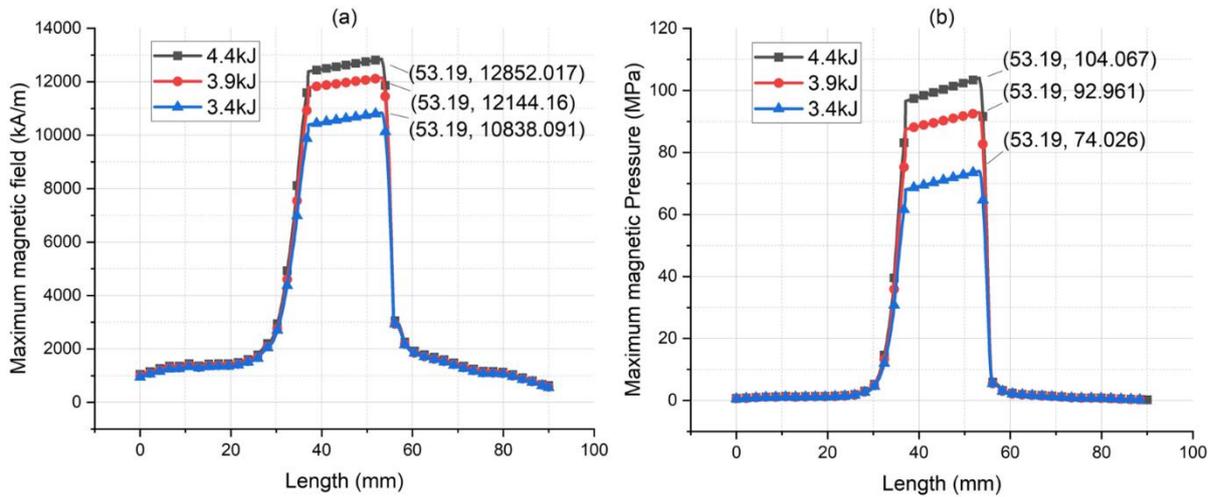

Figure 7 (a) Maximum magnetic field and (b) Maximum magnetic pressure variation at $1.4 \times 10^{-5} s$ along the length

The maximum pressure is observed between 35 mm and 55 mm, which is the crimping zone. The difference between the values at the entry and exit of the crimping zone is due to the influence of copper and SS tubes. At the entry of the crimping zone, we have only Cu tube. In the crimping zone, we have both copper and SS tube and the field shaper, and at the end, only SS tube is present. Cu having higher electrical conductivity than SS produces a secondary magnetic field of higher magnitude, which interferes with the primary magnetic field and hence the resultant magnetic field is lesser at the Cu end. The variation of the magnetic field and the corresponding magnetic pressure with time is plotted at the position of the maximum magnetic field, i.e., at 53.19 mm, as shown in Figure 8 (a) and (b).

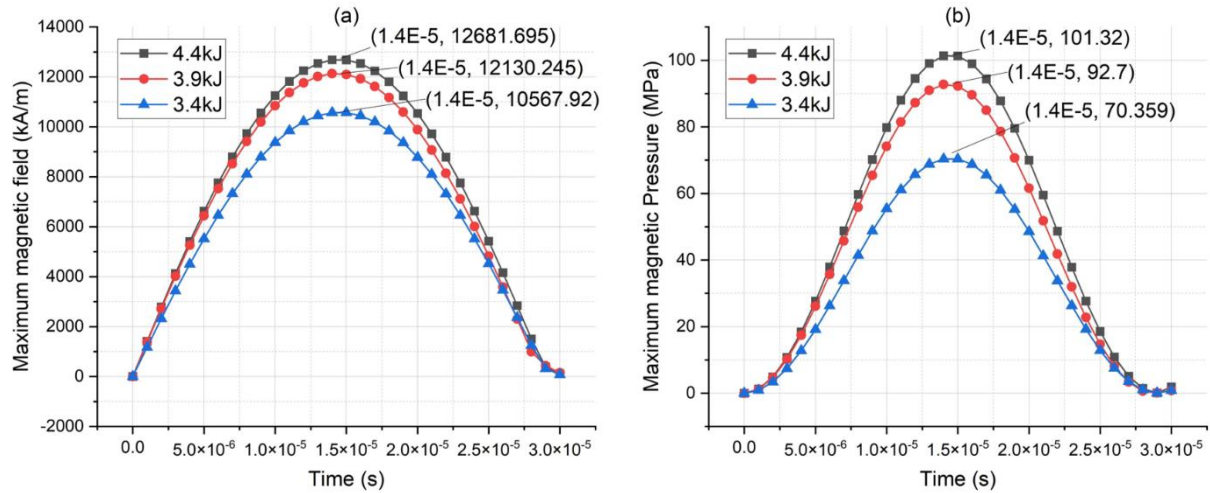

Figure 8 (a) Maximum magnetic field and (b) Maximum magnetic pressure variation

The maximum pressure is observed at time $1.4 \times 10^{-5}$ s. The value of the magnetic field and subsequent pressures increases with time, reaches the maximum value and then decreases.

This time varying pressure data is given as input in ANSYS explicit dynamics.

## 5.3 Convergence Study in ANSYS Explicit Dynamics

In the crimping zone, four nodes are selected along the length of the outer Cu tube at equal intervals as shown in Figure 9. Variation of deformation in these nodes is shown in Figure 10.

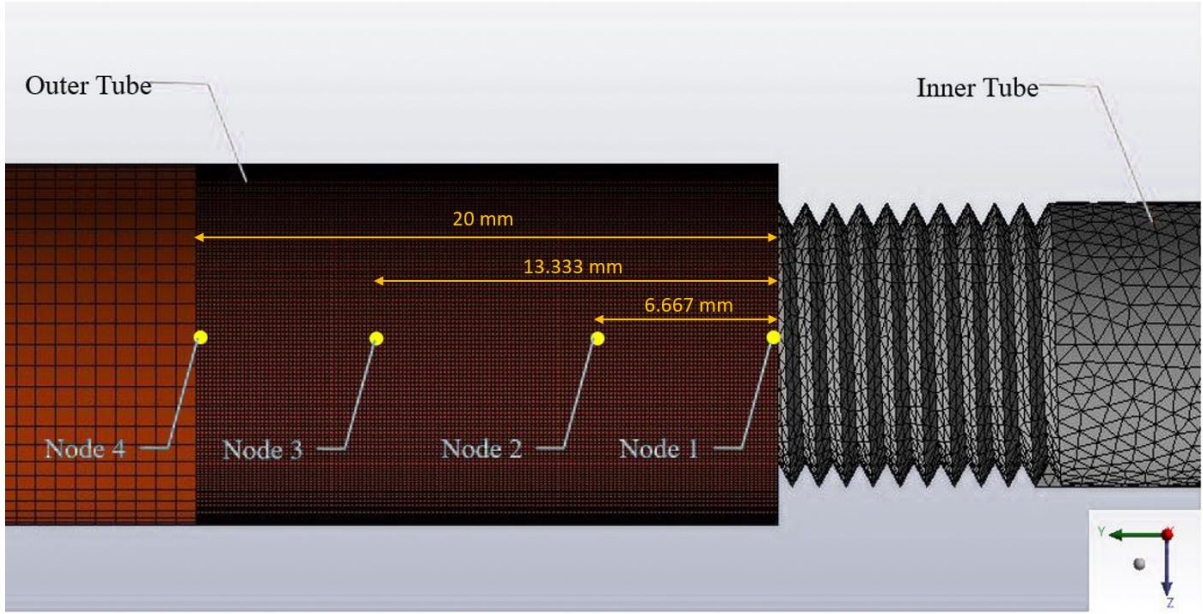

Figure 9 Selected nodes for analysis

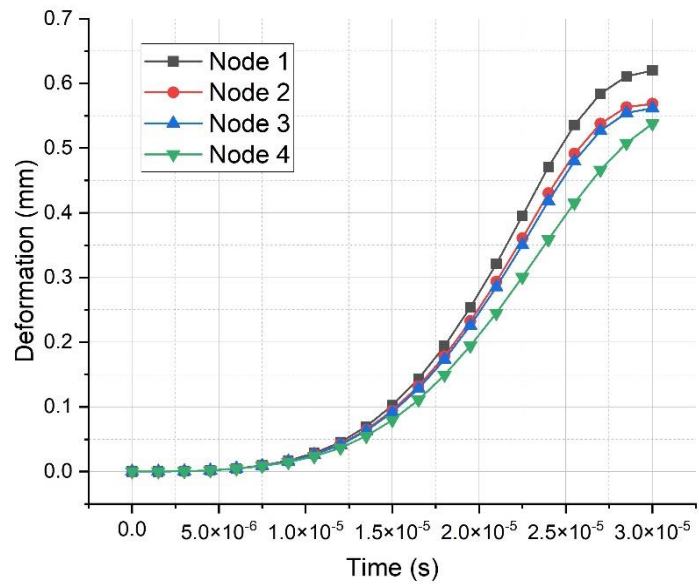

Figure 10 Variation of deformation at the four selected nodes

At a constant time step of $2 \times 10^{-10}$ s, mesh refinement is done with three different mesh sizes 0.2 mm, 0.1 mm and 0.05 mm. Figure 11 shows the stress, strain and deformation at three different mesh sizes. These analyses are done at Node 1 for discharge energy of 4.4 kJ and a pitch of the thread 1.25 mm.

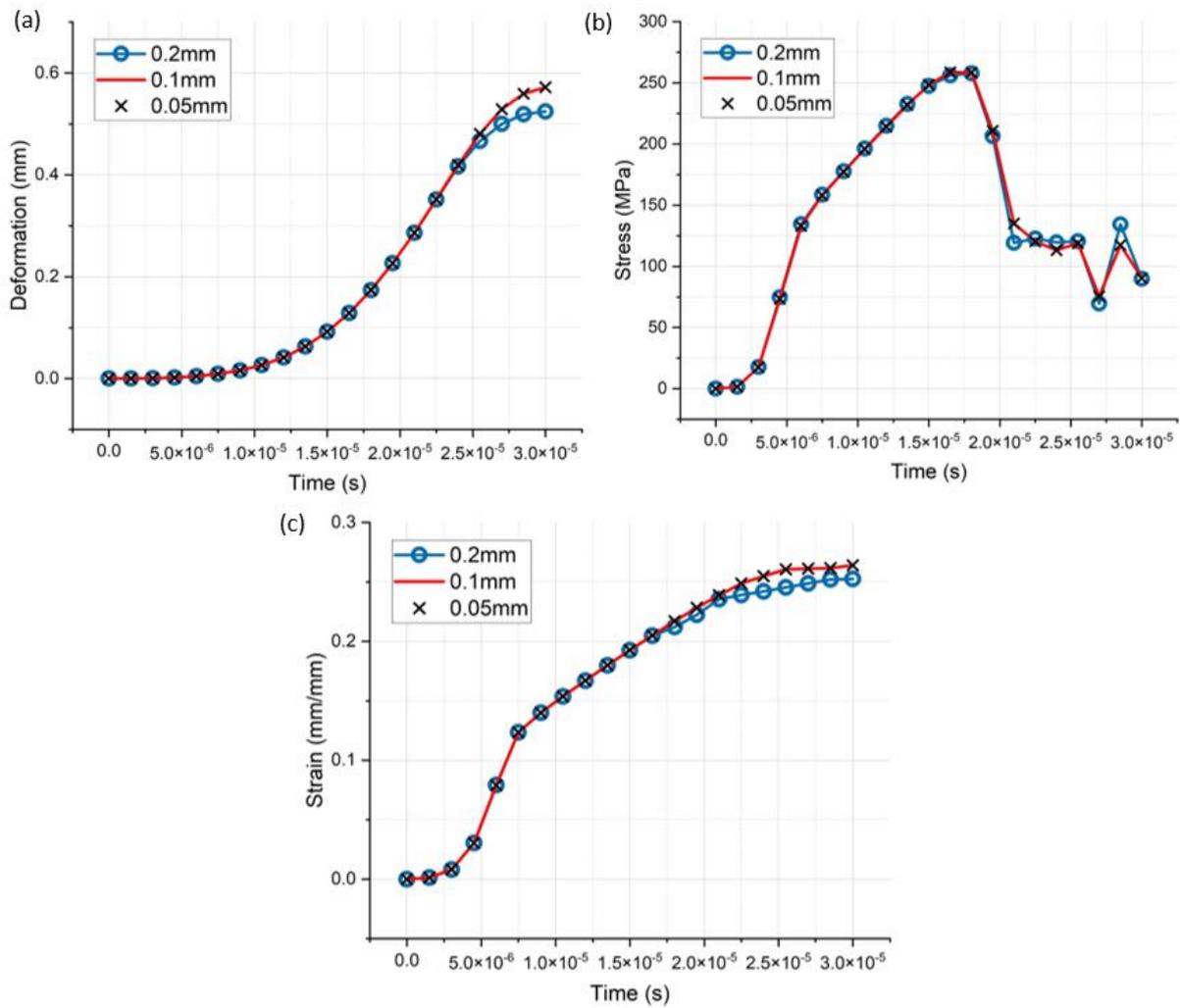

Figure 11 Variation of (a) deformation (b) von Mises stress (c) von Mises strain with time at different mesh sizes

The results are observed to converge at 0.1 mm and 0.05 mm. So, the optimum mesh size is selected as 0.1mm for the subsequent analysis.

At a constant element size of 0.05mm, the time step is varied as $2 \times 10^{-9}$ s, $2 \times 10^{-10}$ s and $1 \times 10^{-10}$ s. Figure 12 shows the stress, strain and deformation, respectively, at Node4, at three different timesteps.

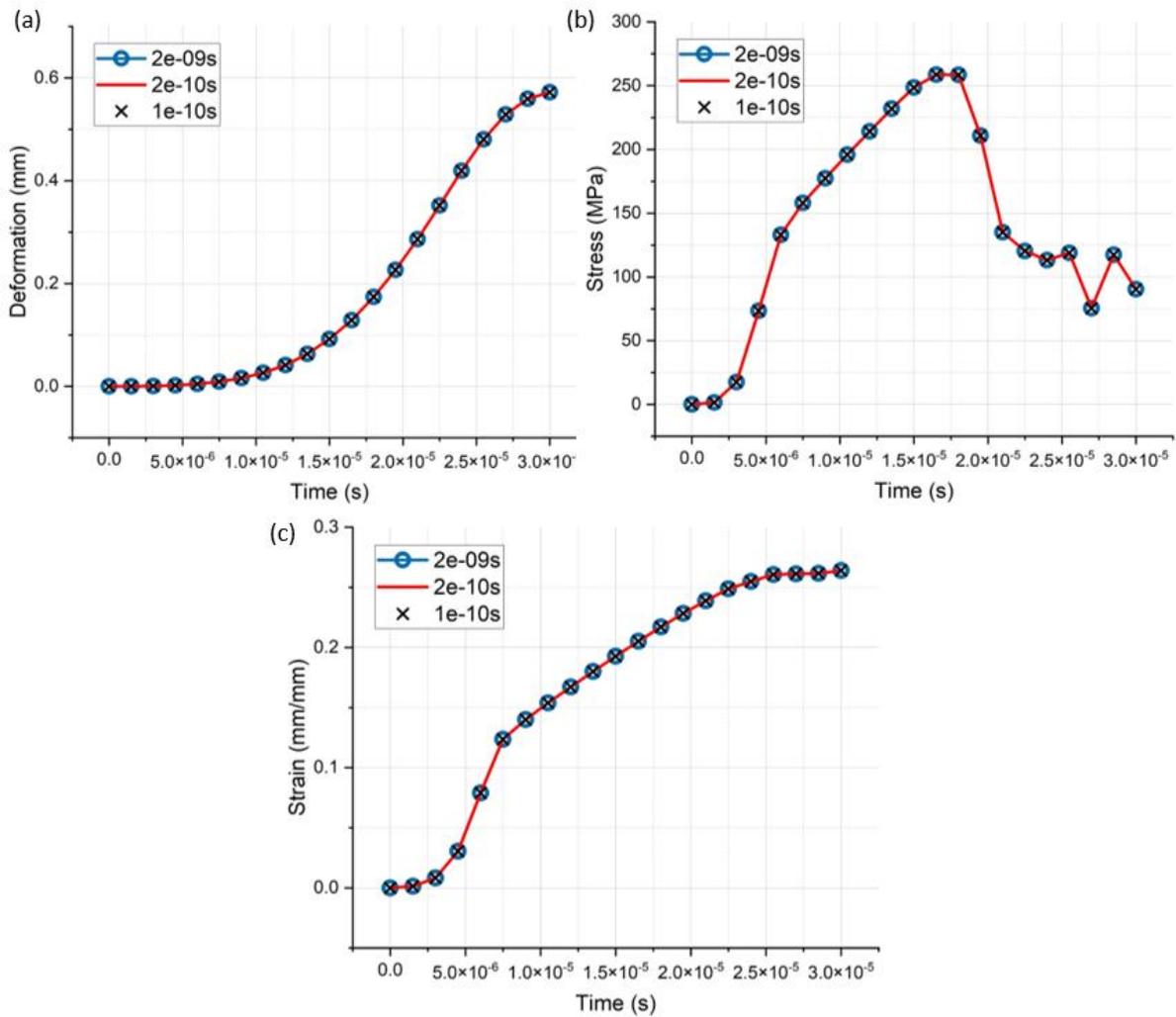

Figure 12 Variation of (a) deformation (b) von Mises stress (c) von Mises strain with time at different timesteps

The results converge for all time steps. So, the time step $2 \times 10^{-10}$s is selected as the optimum time step for the subsequent analysis.

### 5.4 Analysis of Deformation

From Figure 12 (a), it is observed that the maximum value of the stress occurs between 1.5 s and 2 s. The value of stress is above the yield stress of the copper tube so that the tube deforms but less than its ultimate strength so that no failure occurs. Initially, it is observed that the stress keeps on increasing till the deformation starts and then it starts decreasing. In the end, the stress is fluctuating till the completion of the process. The von-Mises strain, as shown in Figure 12 (b) increases very slowly in the beginning. The strain keeps on increasing and reaches the maximum value when the outer tube impacts the inner tube. After that, till the completion of the process, the strain remains almost constant. Deformation is following an increasing trend throughout the time period, as shown in Figure 12 (c) and is maximum at the end time.

Analysis of deformation is done for different combinations of pitches of thread and different discharge energies, i.e., for combinations of discharge energies 3.4 kJ, 3.9 kJ and 4.4 kJ with pitches of the thread 0.75 mm, 1 mm, 1.25 mm and 1.5 mm. The height of the material of the outer tube, which deforms into the thread, is measured as $h_c$, as shown in Figure 13.

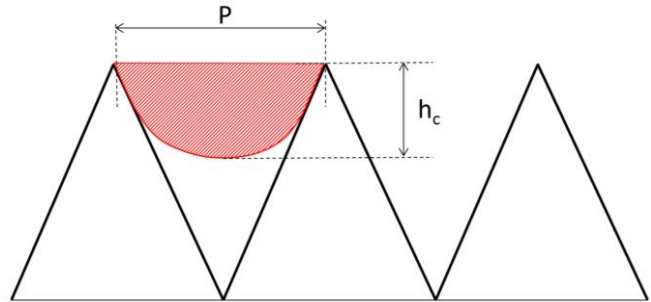

Figure 13 Calculation of $h_c$

With increasing discharge energy, the amount of deformation of the outer tube also increases. The maximum deformation is observed at discharge energy of 4.4 kJ. No permanent deformation is observed in the inner SS tube as it only undergoes elastic deformation during the crimping process. The amount of material of the outer tube entering the thread is also found to be increasing with the pitch of the thread. The deformation is averaged over the crimping zone length to calculate $h_c$.

## 6   Experimental validation

The experiment is conducted for the three discharge energies, and the deformation is measured. In the crimping zone. The outer diameter of the outer tube and its average thickness over length is studied experimentally using an optical microscope with $10 \times$ magnification.

Deformation increases proportionally with the discharge energy as well as with the pitch of the thread. Figure 14 shows the comparison of values obtained from experimental observation and simulation. The experiment values and the simulation values are in perfect agreement for all pitch and discharge energy combinations.

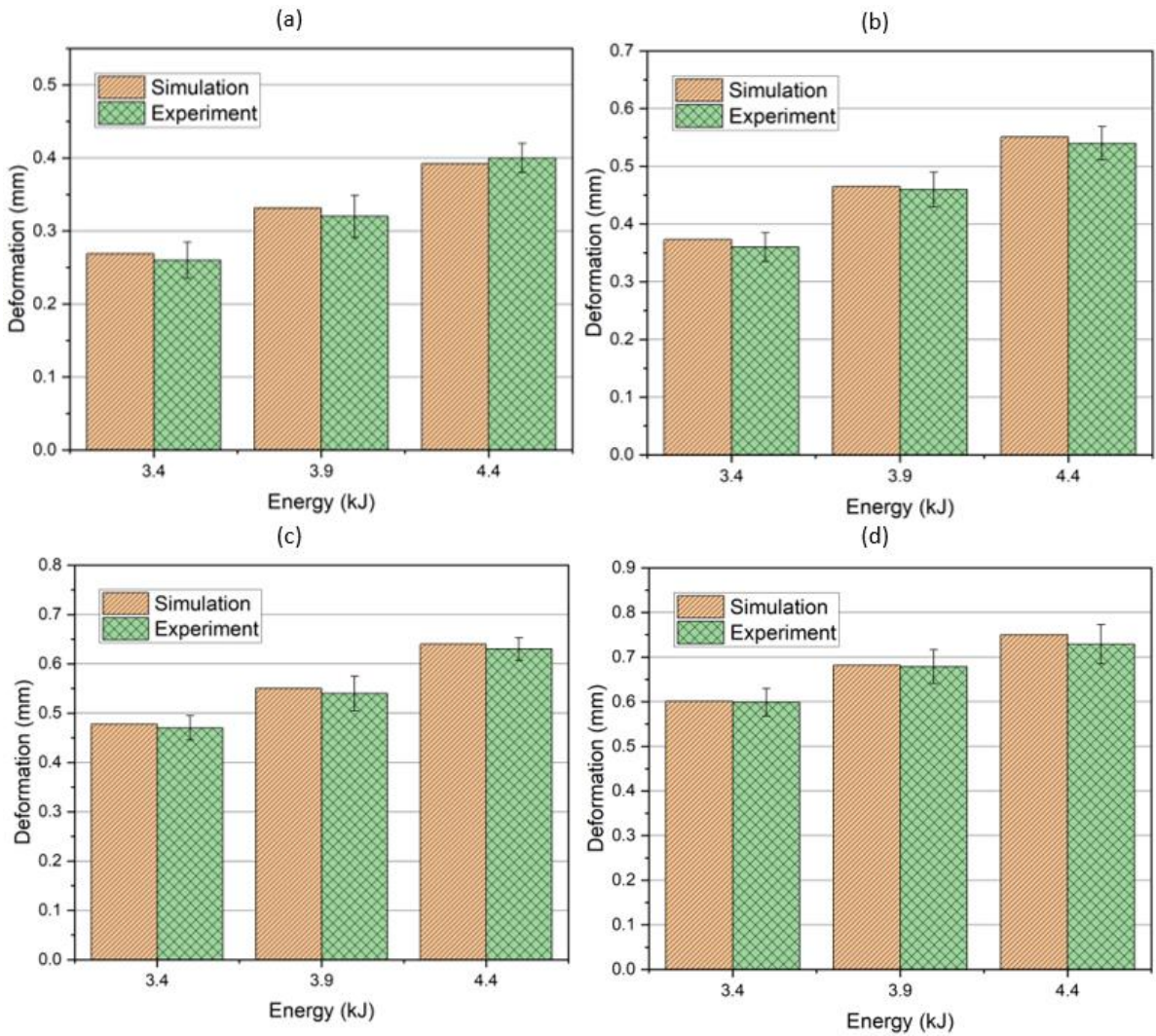

Figure 14 Experiment and Simulation values of $h_c$ for a pitch of thread (a) 0.75mm (b) 1 mm (c) 1.25mm (d) 1.5mm

## 7 Empirical Relation

### 7.1 Deriving Empirical relation from experimental data

Using the experimental data, an empirical relation is developed in MATLAB Curve fitting. Bisquare method, which minimises a weighted sum of squares, is used. The distance of the data point from the fitted line determines the weight of the point. Points closer to the line gets the full weight, and those away from the line gets reduced weight. Zero weight is given to points that are farther from the line than expected. [32]

The bisquare method is usually preferred over the Least Absolute Residuals (LAR) method because while using the usual least-squares approach, it tries to obtain a curve that can fit the majority of the data by minimising the effect of outliers. Rather than the squared difference of the residuals, LAR seeks to find a curve that minimises their absolute differences.

When a simple empirical model is required, polynomials are often preferred. Polynomial fits provide reasonable flexibility for less complicated data, and the fitting process is simple. They do have a disadvantage of unstable high-degree fits. Also, within the data range, they can provide a good fit, but outside that range, they may diverge wildly.

Here, the prediction is made only inside the range of the upper and lower bounds of pitch (0.75-1.5mm) and discharge energies (3.4kJ-44kJ) used in the experiment.

If the deformation is denoted by $h_c$, the pitch is denoted by P, and the discharge energy is denoted by $E$, the empirical relation can be expressed in the form,

$$h_c = a_{00} + a_{10}P + a_{01}E + a_{20}P^2 + a_{11}PE + a_{02}E^2 + a_{30}P^3 + a_{21}P^2E + a_{12}PE^2 \tag{13}$$

The coefficients (with 95% confidence bounds) derived using the bisquare method are shown in Table 4.

Table 4 Coefficients of empirical relation fitted using bisquare method

| $a_{00}$ | $a_{01}$ | $a_{02}$ | $a_{10}$ | $a_{11}$ | $a_{12}$ | $a_{20}$ | $a_{21}$ | $a_{30}$ |
|---|---|---|---|---|---|---|---|---|
| 1.168 | -0.8989 | 0.09493 | -0.7936 | 1.306 | -0.0889 | -1.17 | -0.2807 | 0.6659 |

The curve fitted using the Bisquare method is shown in Figure 15. The black dots represent the experimental data points.

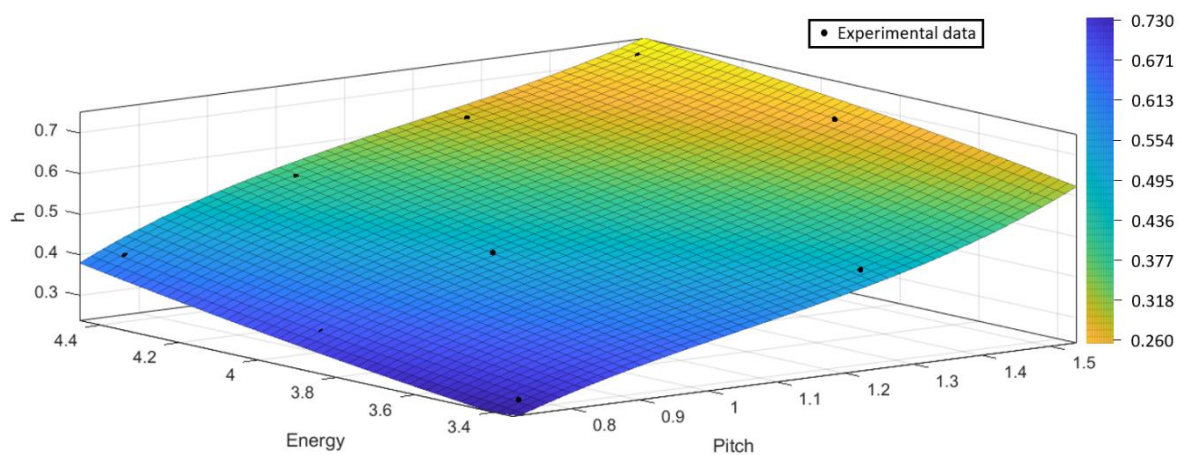

Figure 15 Obtained solution surface through curve fitting using bisquare method

The coefficients (with 95% confidence bounds) derived using LAR method are shown in Table 5. The curve fitted using the LAR method is shown in Figure 16. The black dots represent the experimental data points.

Table 5 Coefficients of empirical relation fitted using LAR method

| $a_{00}$ | $a_{01}$ | $a_{02}$ | $a_{10}$ | $a_{11}$ | $a_{12}$ | $a_{20}$ | $a_{21}$ | $a_{30}$ |
|---|---|---|---|---|---|---|---|---|
| 1.947 | -1.265 | 0.14 | -1.673 | 1.673 | -0.133 | -1.023 | -0.2874 | 0.6293 |

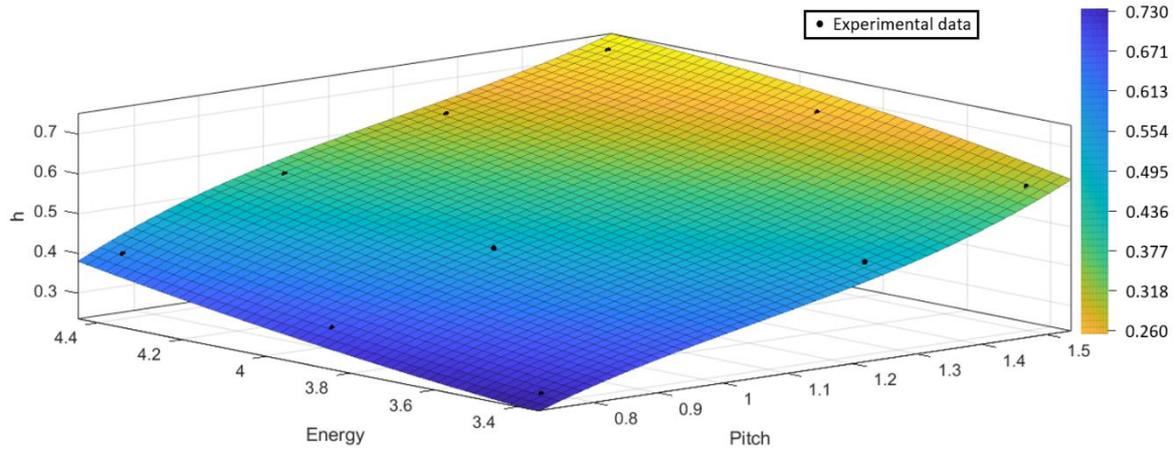

Figure 16 Obtained solution surface through curve fitting using LAR method

### 7.2 Validation of empirical relation with simulation

The L$_2$ error norm of the developed relations are calculated using the equation:

$$\mathrm{L_2 error} = \frac{\sum_{i=1}^{N}\left(h_{c_{s_i}} - h_{c_{e_i}}\right)^2}{\sum_{i=1}^{N} h_{c_{e_i}}^2} \qquad (14)$$

where i denotes a particular combination of discharge energy and pitch, $h_{c_{s_i}}$ is $h_c$ from simulation for i$^{\text{th}}$ case, $h_{c_{e_i}}$ is $h_c$ from empirical relation for i$^{\text{th}}$ case.

The $L_2$ error norm for the Bisquare method is calculated to be $8..77 \times 10^{-5}$ and the $L_2$ error norm for the LAR method is calculated to be $1.2074 \times 10^{-4}$. This error is calculated over the 30 combinations of pitch and discharge energies, i.e. for combinations of discharge energies 3.4kJ, 3.65kJ, 3.9kJ, 4.15kJ, and 4.4kJ with pitches of the thread 0.75mm, 1mm, 1.15mm, 1.25mm, 1.4mm and 1.5mm.

The deformation predicted by the empirical relations is validated using simulations for the above combination of six pitches and five discharge energies, including two new intermediate pitches (1.15mm and 1.4mm) and two new intermediate discharge energies (3.65kJ and 4.15kJ). Bisquare method is chosen for this as it has a lesser error norm. Figure 17 shows the variation of $h_c$ with discharge energy for different pitches, and Figure 18 shows the variation of $h_c$ with the pitch of the thread for different discharge energies. The markers represent the simulation data, and the lines represent the predicted values by the empirical relation. Both Figure 17 and Figure 18 represent a perfect match between the empirical relation derived from experimental data and the numerical simulations using ANSYS Maxwell and ANSYS explicit dynamics. There is also a very consistent match between empirically predicted deformation and numerically simulated deformations for the discharge energies 3.65 kJ and 4.15 kJ, and pitches 1.15 mm and 1.4 mm. These data sets are not used while deriving the empirical relations from experimental results. Therefore, the empirical relation is proven to be successful in predicting $h_c$ within the range of discharge energies.

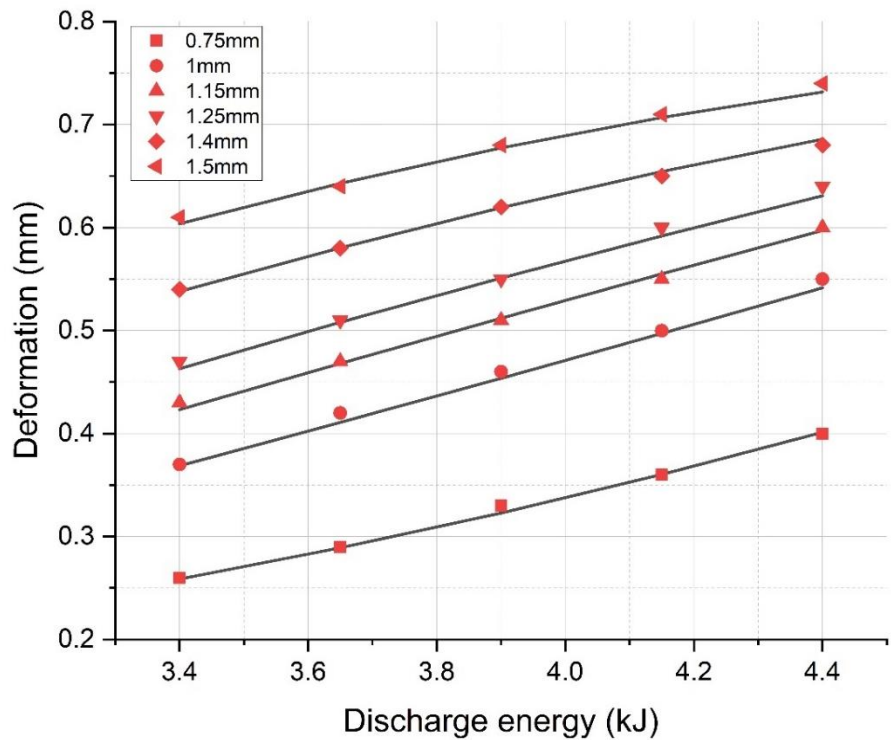

Figure 17 $h_c$ vs Discharge energy for different pitches

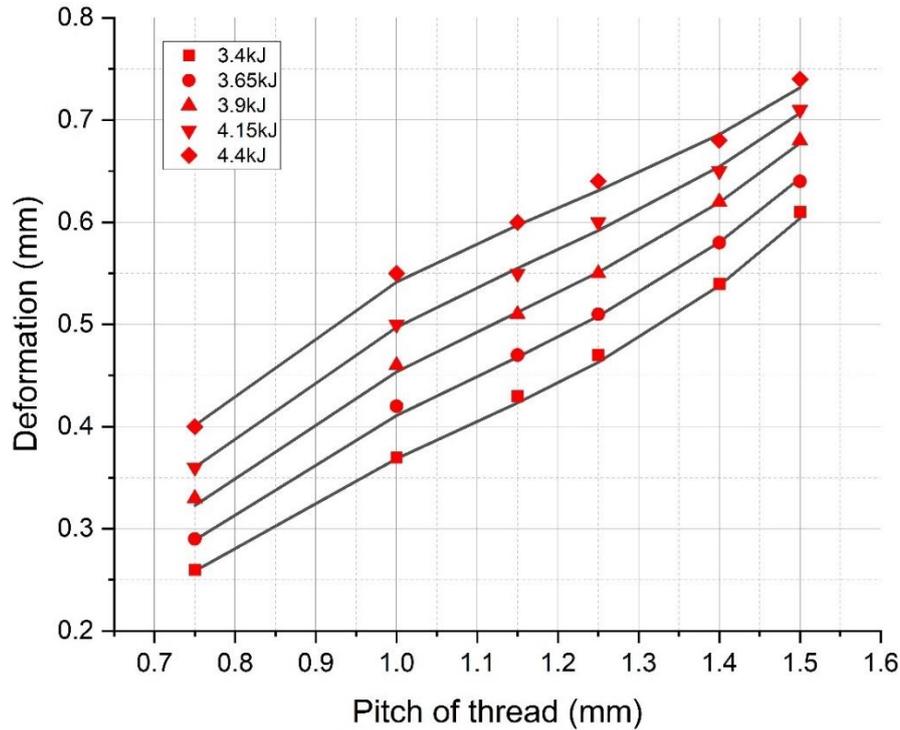

Figure 18 $h_c$ vs Pitch of thread for different discharge energies

## 8 Conclusion

Finite element non-coupled modelling of electromagnetic crimping of Copper-Stainless steel with threaded stainless-steel inner tube is presented in the current article using ANSYS Maxwell and ANSYS Explicit Dynamics.

- Numerical simulations are done for multiple combinations of discharge energies and pitches of the thread.
- Deformation, stress, strain of the tubes are analysed.
- Experiments are conducted in an electromagnetic forming machine to validate the numerical results. The deformation of the tube calculated by the simulations agrees with that in the experiments.
- The amount of deformation for a particular pitch of the thread is observed to be increasing with the discharge energy.
- For a particular discharge energy, the deformation is increasing with the pitch of the thread.
- An empirical relation is developed from the experimental data by curve fitting using the bisquare method and LAR method. Bisquare method has a lesser $L_2$ error norm, so it is chosen as the predicting relation

- The relation successfully predicts the deformation inside the range of discharge energies and pitches of the thread from the experiment. These empirically obtained deformations also closely match with the FEM simulated deformations.
- The empirically predicted deformation values are further validated with FEM simulations for intermediate pitches and discharge energies which have not been considered in deriving the empirical relation.

**Acknoledgment**

The authors are grateful to the SERB, DST, India for supporting this research under Project (IMP/2019/000276), for financial support through fast-track young scientist scheme to carry out research work and for Research scholarship from MHRD, India